\documentclass[english,twocolumn,twocolumn,showpacs,preprintnumbers,amsmath,amssymb]{revtex4}
\usepackage[T1]{fontenc}
\usepackage[latin1]{inputenc}
\usepackage{graphicx}

\makeatletter

\makeatletter

\makeatletter

\usepackage{dcolumn}

\usepackage{bm}

\newcommand{\pt}{ p_{\rm T}}

\newcommand{\eg}{{\it e.g.}}
\newcommand{\ie}{{\it i.e.}}
\newcommand{\beq}{\begin{equation}}
\newcommand{\eeq}{\end{equation}}
\newcommand{\bea}{\begin{eqnarray}}
\newcommand{\eea}{\end{eqnarray}}
\newcommand{\bef}{\begin{figure}}
\newcommand{\eef}{\end{figure}}
\newcommand{\bce}{\begin{center}}
\newcommand{\ece}{\end{center}}

\def\lsim{\mathrel{\rlap{\lower4pt\hbox{\hskip1pt$\sim$}}
    \raise1pt\hbox{$<$}}}         %less than or approx. symbol
\def\gsim{\mathrel{\rlap{\lower4pt\hbox{\hskip1pt$\sim$}}
    \raise1pt\hbox{$>$}}}         %greater than or approx. symbol

\makeatother

\makeatother

\usepackage{babel}
\makeatother
\begin{document}
%\preprint{APS/123-QED}

\title{A Systematic Study on Direct Photon Production from Central Heavy
Ion Collisions }

\author{Fu-Ming Liu}

\email{liufm@iopp.ccnu.edu.cn}

\affiliation{Institute of Particle Physics, Central China Normal University, Wuhan,
China }

\author{Klaus Werner}

\affiliation{Laboratoire SUBATECH, University of Nantes - IN2P3/CNRS - Ecole des
Mines, Nantes, France }

\date{\today}

\begin{abstract}
We investigate the production of direct photons in central Au-Au collisions
at the relativistic Heavy-Ion Collider (RHIC) at 200~GeV per nucleon,
considering all possible sources. We treat thermal photons emitted
from a quark-gluon plasma and from a hadron gas, based on a realistic
thermodynamic expansion. Hard photons from elementary nucleon-nucleon
scatterings are included: primordial elementary scatterings are certainly
dominant at large transverse momenta, but also secondary photons from
jet fragmentation and jet-photon conversion cannot be ignored. In
both cases we study the effect of energy loss, and we also consider
photons emitted from bremsstrahlung gluons via fragmentation. 
\end{abstract}
\pacs{25.75.-q,12.38.Mh}
%\keywords{}
\maketitle

\section{Introduction}

The formation and observation of a quark-gluon plasma in heavy ion
collisions is an important goal of modern nuclear physics. Even if
a quasiequilibrated plasma is created for a brief time in the collisions,
it is still a challenge to infer knowledge of the plasma from particle
production. Among those proposed \char`\"{}probes\char`\"{} of the
plasma are the directly produced real photons. Those photons interact
only electromagnetically, and so their mean free paths are typically
much larger than the transverse size of the hot dense matter created
in the collisions. As a result, high-energy photons produced in the
interior of the plasma usually pass through the surrounding matter
without interaction, carrying information directly from wherever they
were formed to the detector.

One expects to deduce properties of the highly excited matter, in
particular it's space-time evolution. Direct photons may provide these
informations in two different ways: via so-called \char`\"{}thermal
photons\char`\"{} emitted directly from hot and dense matter, and
via secondary photons from initially produced jets. There are two
kinds of such secondary photons: they may originate from the \char`\"{}fragmentation\char`\"{}
of jets, or the \char`\"{}conversion\char`\"{} of jets into photons
via interactions of jets with the partons of the equilibrated matter.
Such secondary photons compete, however, with the photons directly
produced in an initial hard nucleon-nucleon scattering, in the following
referred to as \char`\"{}primordial NN scattering\char`\"{} contribution,
so a carefully study of these photons is very important. In the following,
we will shortly review these different photon sources.

\textbf{\textit{Primordial NN scattering.}} The direct photon production
via Compton scattering and quark-antiquark annihilation can be calculated
in perturbation theory, using the usual parton distribution functions.
In principle one should consider at this stage also higher order contributions,
like bremsstrahlung of photons accompanying for example two-jet production
in hards parton-parton scattering. However, we consider this as part
of the so-called jet fragmentation (or bremsstrahlung) contribution,
which will be affected by the thermalize matter, and which we will
discuss separately.

\textbf{\textit{Thermal photons.}} In high energy nuclear collisions,
the density of secondary partons is so high that the quarks and gluons
rescatter and eventually thermalize to form a bubble of hot quark-gluon
plasma (QGP).The plasma expands and decreases its energy density so
that a phase transition to hadronic gas (HG) phase appears. Thermal
photons can be produced during the whole history of the evolution,
from the QGP phase, the mixed phase, and from the pure HG phase. Photons
from a thermal source are exponentially damped so that the contribution
to very high $\pt$ is negligible. However, its contribution to low
$\pt$ is dominant.

\textbf{\textit{Jet-photon conversion.}} When jets pass through thermalized
matter, they may interact. In case of the quark-gluon plasma, these
interactions are elastic collisions between jets and deconfined partons.
There is first of all quark-antiquark annihilation and quark-gluon
Compton scattering, which both can produce a photon. These photons
are affected by the plasma in two ways: obviously the plasma is needed
to allow these interactions, but there is also a secondary effect,
since the jets first of all lose energy in the plasma, before contributing
to the photon production.

\textbf{\textit{Jet fragmentation or bremsstrahlung photons.}} Photon
production also occurs as higher order effect in purely partonic initial
hard scatterings: at any stage of the evolution of a jet (final state
parton emissions), there is a possibility of emitting photons. Also
here the presence of a QGP will affect the results, since the jets
lose energy during the fragmentation process. And not to forget the
fragmentation contribution from induced gluon radiation in the plasma.

Our paper is organized as follows: in sections 2 and 3, we discuss
thermal photon production; in section 4, we study hard photons from
primordial NN scattering (leading order); in section 5 we study jet-photon
conversion, including the modification due to jet energy loss; in
section 6 we compute photon production from jet fragmentation, also
referred to as bremsstrahlung's photons, again considering the effect
of energy loss; in section 7 we finally collect our results and compare
with experimental data.

\section{Photon emission rates}

Thermal photon production is obtained by integrating the photon emission
rate $R$ (number of reactions per unit time per unit volume which
produce a photon ) over the space-time history of the expanding hot
and dense matter. In this section we study the photon emission rates
from different phases of the hot dense matter.

The spectral photon emissivity directly reflects the dynamics of real
photon production reactions in thermalized matter. Commonly employed
formalisms are finite-temperature field theory and kinetic theory.
As systematically studied by Kapusta \textit{et al.}\cite{Kapusta1991},
the thermal emission rate of photons with energy $E$ and momentum
$\vec{p}$ from a small system (compared to the photon mean free path)
is \begin{equation}
E\frac{dR}{d^{3}p}=\frac{-2}{(2\pi)^{3}}{\rm Im}\Pi_{\mu}^{R,\mu}\frac{1}{\exp(E/T)-1}\label{eq:finite-temp. filed theory}\end{equation}
 where $\Pi_{\mu}^{R,\mu}$ is the retarded photon self-energy at
finite temperature $T$. This formula has been derived both perturbatively
and nonperturbatively. It is valid to all orders in the strong interaction.
If the photon self-energy is approximated by carrying out a loop expansion
to some finite order, then the formulation of Eq. (\ref{eq:finite-temp. filed theory})
is equivalent to relativistic kinetic theory, where the emission rate
of photons with energy $E$ and momentum $\vec{p}$ from a process
of type 1+2$\to$3+$\gamma$ reads\begin{eqnarray}
E\frac{dR}{d^{3}p} & = & \int(\prod_{i=1}^{3}\frac{d^{3}p_{i}}{(2\pi)^{3}2E_{i}})(2\pi)^{4}\delta^{(4)}(p_{1}^{\mu}+p_{2}^{\mu}-p_{3}^{\mu}-p_{\gamma}^{\mu})\nonumber \\
 &  & \times\left|{\cal M}\right|^{2}\frac{f_{1}(E_{1})f_{2}(E_{2})[1\pm f_{3}(E_{3})]}{2(2\pi)^{3}}\label{eq:kinet}\end{eqnarray}
 where the $f$'s are the Fermi-Dirac or Bose-Einstein distribution
functions as appropriate. Eq.(\ref{eq:kinet}) is convenient if the
scattering amplitude, $M$, is evaluated in a perturbative expansion.
Non-perturbative (model) calculations at low and intermediate energies,
on the other hand, are more amenable to the correlator formulation,
Eq.(\ref{eq:finite-temp. filed theory}). In the hadronic medium,
$\eg$, $\Pi_{em}$ can be directly related to vector-meson spectral
functions within the vector dominance model (VDM). Instructive investigation
on photon emission rates from both QGP and HG phases can be found
in \cite{Kapusta1991}\cite{Rapp_rev_2004}.

\subsection{Quark-Gluon Plasma}

\begin{figure}
\includegraphics[scale=0.35]{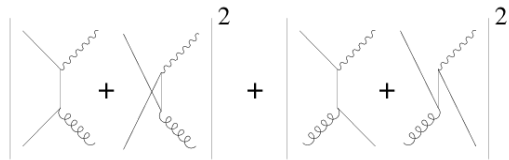}

\caption{\label{fig:FDg_QGP} Two-to-two particle processes contributing to
the leading order photon emission rate. }
\end{figure}

In \cite{Kapusta1991}, the thermal rate from a quark gluon plasma
is computed using the kinetic theory formalism for the simplest two-to-two
scattering diagrams such as the QCD Compton process $qg\rightarrow\gamma q$
and annihilation $q\bar{q}\rightarrow g\gamma$, see Fig.\ref{fig:FDg_QGP}.
In case of a large photon energy, and with the energy of two initial
partons being larger than the energy of the output photon ($E_{1}+E_{2}>E\gg T$),
the approximation $f_{1}(E_{1})f_{2}(E_{2})\rightarrow\exp[-(E_{1}+E_{2})/T]$
is employed. Because the light quark masses are set to zero, an infrared
cutoff $-k_{c}^{2}$ must be placed on the four-momentum transfer.
The infrared divergence is regulated by an infinite resummation of
finite-temperature Feynman diagrams, following Braaten and Pisarski.
This amounts to a careful treatment of a small part of phase space
left out in the kinetic theory calculation by imposing the infrared
cutoff. When the contributions from the two regions of phase space
-- below and above the cutoff -- are added, the result is independent
of the cutoff. This provides a parameterized thermal emission rate
of photons with energy $E$ and momentum $\vec{p}$ from an equilibrated
QGP at temperature $T$ and zero net baryon density, for large values
of $x=E/T$, given as \begin{equation}
E\frac{dR^{\mathrm{QGP}\rightarrow\gamma}}{d^{3}p}=\sum_{i=1}^{N_{f}}(\frac{e_{i}}{e})^{2}\frac{\alpha\alpha_{S}}{2\pi^{2}}T^{2}{\rm e}^{-x}\ln\left(1+\frac{2.912}{4\pi\alpha_{s}}x\right)\ \label{eq:rate_pert}\end{equation}
 An additive {}``1\char`\"{} has been introduced in the argument
of the logarithm to enable extrapolation to small $x$ \cite{Kapusta1991}.

\begin{figure}
\includegraphics[scale=0.25]{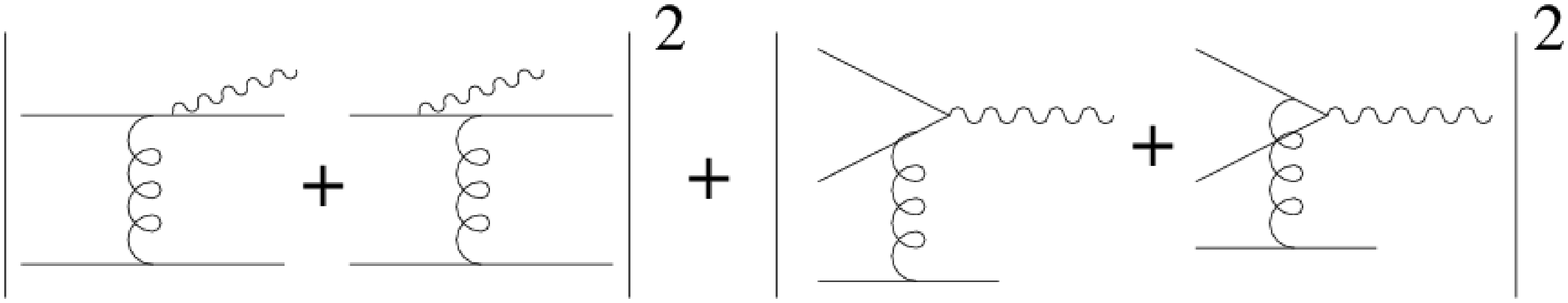}

\caption{\label{fig:FDg_QGP2} Bremsstrahlung and pair production contributing
to photon emission. The bottom line in each diagram can represent
either a quark or a gluon.}
\end{figure}

\begin{figure}
\includegraphics[scale=0.35]{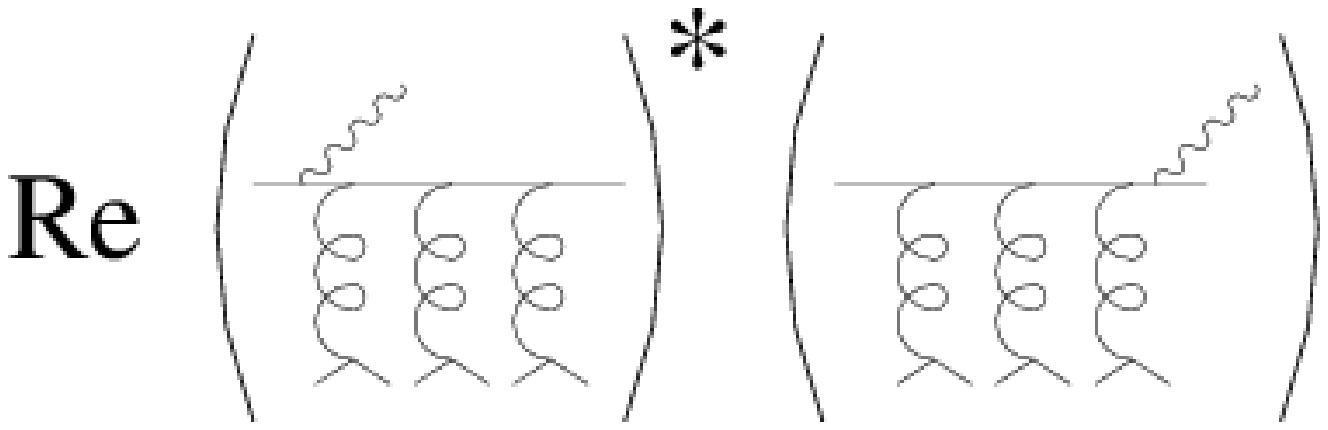}

\caption{\label{fig:FDg_QGP3} An interference term, involving amplitudes
for photon emission before and after multiple scattering events, which
contributes to the leading order emission rate. }
\end{figure}

As noticed in Ref.\cite{Arenche_gelis} , Eq.(\ref{eq:rate_pert})
does not yet comprise the full result to leading order in the strong
coupling constant $\alpha_{s}$. Due to collinear singularities, bremsstrahlung
as well as pair annihilation graphs contribute at the same order as
the resumed 2$\to$2 processes, c.f. Fig.\ref{fig:FDg_QGP2}. The
full result, which also necessitates the incorporation of Landau-Pomeranchuk-Migdal
(LPM) interference effects, as shown in Fig.\ref{fig:FDg_QGP3}, has
been computed in Ref.\cite{AMY2001} as \begin{eqnarray}
E\frac{dR^{\mathrm{QGP}\rightarrow\gamma}}{d^{3}p} & = & \sum_{i=1}^{N_{f}}(\frac{e_{i}}{e})^{2}\frac{\alpha\alpha_{S}}{2\pi^{2}}T^{2}\frac{1}{e^{x}+1}\label{eq:rate_lo}\\
 &  & \times\big[\ln(\frac{\sqrt{3}}{g})+\frac{1}{2}\ln(2x)+C_{22}(x)\nonumber \\
 &  & \qquad+C_{\mathrm{brems}}(x)+C_{\mathrm{ann}}(x)\big]\nonumber \end{eqnarray}
 with convenient parameterizations of the 3 functions $C$ as \cite{AMY2001}
\begin{equation}
C_{22}(x)=\frac{0.041}{x}-0.3615+1.01{\rm e}^{-1.35x}\qquad\qquad\label{eq:c22}\end{equation}
 \begin{eqnarray}
 &  & C_{\mathrm{brems}}(x)+C_{\mathrm{ann}}(x)\label{eq:2cs}\\
 &  & \qquad=0.633x^{-1.5}\ln\left(12.28+1/x\right)+\frac{0.154x}{(1+x/16.27)^{0.5}}\ .\nonumber \end{eqnarray}
 We will employ the above formula, taking $N_{f}=3$, and a temperature
dependent running coupling constant\cite{Karsch1998} \begin{equation}
\alpha_{s}(T)=\frac{6\pi}{(33-2N_{f})\ln(8T/T_{c})}.\label{eq:alphasT}\end{equation}
 Effects from non-zero baryon density and from off-equilibrium are
not included, the above rates can be considered to be a good approximation
for the midrapidity region of Au+Au collisions at 200~AGeV. In Fig.\ref{fig:Rate1}
we show photon emission rate from QGP. The dashed line represents
the results obtained by Kapusta \textit{et al}\cite{Kapusta1991},
summarized in eq.(\ref{eq:rate_pert}), the solid line refers to the
calculations of AMY\cite{AMY2001}, given in eq.(\ref{eq:rate_lo}),
with additional contributions compared to \cite{Kapusta1991}. We
can see the full contribution to photon emission from a QGP is much
higher than the one from $2\rightarrow2$ partonic processes.

\subsection{Hadronic Matter}

\begin{figure}
\includegraphics[scale=0.35]{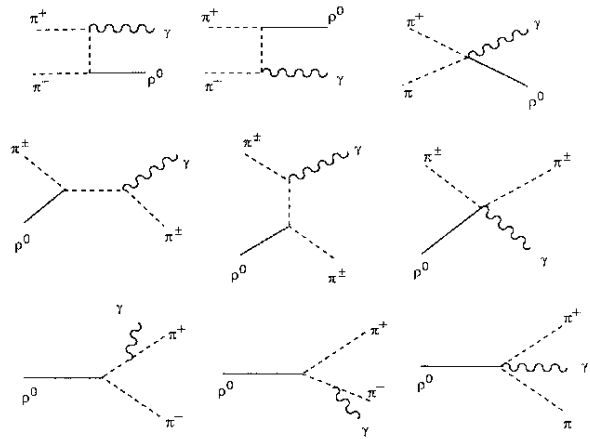}

\caption{\label{fig:FDg_HG1} Photon production reactions and decays involving
charged pions (dashed lines) and neutral $\rho$ mesons (solid lines).}
\end{figure}

Photons can also be produced in a hadronic phase, from several elementary
interactions, see Fig.\ref{fig:FDg_HG1}. The dominant contribution
comes from the reactions $\pi\pi\rightarrow\rho\gamma$ and $\pi\rho\rightarrow\pi\gamma$;
The decay $\rho\rightarrow\pi+\pi+\gamma$ also contributes significantly.
Interactions involving strange mesons or baryons can also produce
photons, but these contributions are relatively small because of the
phase space suppression due to their big masses. The situation of
thermal photon radiation rates from a hadronic gas is uncertain, due
to difficulties related to the strong coupling and the masses of hadrons.
The study is usually carried out within effective Lagrangians. Constraints
on the interaction vertices can, to a certain extent, be imposed by
symmetry principles ($\eg$, e.m. gauge and chiral invariance). Coupling
constants are estimated by adjusting to measured decay branchings
in the vacuum. Thus, for the temperature ranges relevant to practical
applications, $T$=100-200~MeV, the predicted emission rates are
inevitably beset with significant uncertainties, and therefore a careful
judgment of the latter becomes mandatory. %

\begin{figure}
\includegraphics[scale=0.3]{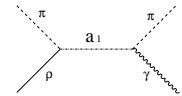}

\caption{\label{fig:FDg_HG2} Feynman diagram of $\pi\rho\to\pi\gamma$ through
$a_{1}(1260)$ resonance. }
\end{figure}

Investigations along these lines were initiated in Ref.\cite{Kapusta1991},
where the photon self-energy has been computed to 2-loop order for
a mesonic system consisting of sharp (zero width) $\pi$-, $\eta$-
and $\rho$-mesons (plus direct $\omega\to\pi^{0}\gamma$ decays).
In Ref.\cite{Xiong1992} it was pointed out that $\pi\rho\to\pi\gamma$
scattering via $a_{1}(1260)$ resonance formation (or, equivalently,
$a_{1}\to\pi\gamma$ decay), c.f. Fig.\ref{fig:FDg_HG2}, constitutes
an important contribution. This was followed up by a systematic treatment\cite{Song1993}
of an interacting $\pi\rho a_{1}$ system to 2-loop order within the
Massive Yang-Mills (MYM) framework of introducing axial-/vector mesons
into a chiral Lagrangian, and, later, within the Hidden-Local Symmetry
(HLS) approach\cite{Hal98}. The effect on in-medium vector and axial-vector
meson masses is studied by Song and Fai \cite{Song1998}. %

\begin{figure*}
\includegraphics[scale=0.75]{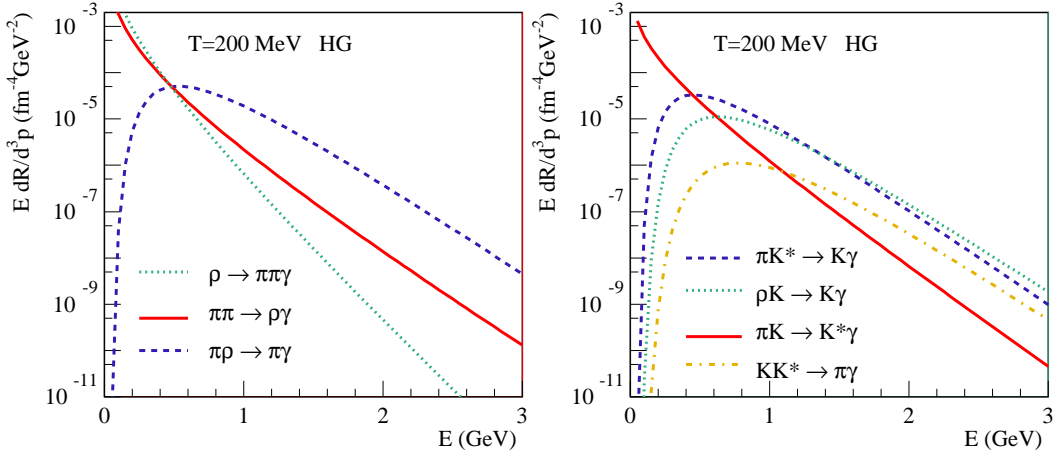}

\caption{\label{fig:Rate2} Photon emission rates from a HG from the different
reactions (bare graphs, no hadronic form factor). }
\end{figure*}

We will use the results of the MYM calculation \cite{Rapp2004},
where photon production from strangeness bearing mesons has been included
as well as the axial meson $a_{1}$ as exchange particle for non-strange
initial states. We show these rates in fig.\ref{fig:Rate2}, and we
list the corresponding parameterized rates, in unit of $\mbox{fm}^{-4}\mbox{GeV}^{-2}$
, with the photon energy ($E$) and the temperature ($T$) both in
GeV:

\begin{eqnarray}
 &  & E\frac{dR_{\pi+\rho\rightarrow\pi+\gamma}}{d^{3}p}=T^{2.8}exp\big(\frac{-(1.461T^{2.3094}+0.727)}{(2TE)^{0.86}}\nonumber \\
 &  & \qquad+(0.566T^{1.4094}-0.9957)\frac{E}{T}\big)\ \end{eqnarray}

\begin{eqnarray}
 &  & E\frac{dR_{\pi+\pi\rightarrow\rho+\gamma}}{d^{3}p}=\frac{1}{T^{5}}exp\big(-(9.314T^{-0.584}\label{eq:pipi}\\
 &  & \qquad-5.328)(2TE)^{0.088}+(0.3189T^{0.721}-0.8998)\frac{E}{T}\nonumber \end{eqnarray}

\begin{eqnarray}
 &  & E\frac{dR_{\rho\rightarrow\pi+\pi+\gamma}}{d^{3}p}=\label{eq:rho}\\
 &  & \qquad\frac{1}{T^{2}}exp\big(-\frac{(-35.459T^{1.126}+18.827)}{(2TE)^{(-1.44T^{0.142}+0.9996)}}-1.21\frac{E}{T}\big)\nonumber \end{eqnarray}

\begin{eqnarray}
 &  & E\frac{dR_{\pi+K^{*}\rightarrow K+\gamma}}{d^{3}p}=T^{3.75}exp\big(-\frac{0.35}{(2TE)^{1.05}}\qquad\qquad\label{eq:pik*}\\
 &  & \qquad\qquad+(2.3894T^{0.03435}-3.222)\frac{E}{T}\big)\nonumber \end{eqnarray}

\begin{eqnarray}
 &  & E\frac{dR_{\pi+K\rightarrow K^{*}+\gamma}}{d^{3}p}=\frac{1}{T^{3}}exp\big(-(5.4018T^{-0.6864}\qquad\qquad\nonumber \\
 &  & \qquad\qquad\qquad-1.51)(2TE)^{0.07}-0.91\frac{E}{T}\big)\label{eq:piK}\end{eqnarray}

\begin{eqnarray}
 &  & E\frac{dR_{\rho+K\rightarrow K+\gamma}}{d^{3}p}=T^{3.5}exp\big(-\frac{(0.9386T^{1.551}+0.634)}{(2TE)^{1.01}}\nonumber \\
 &  & \qquad\qquad+(0.568T^{0.5397}-1.164)\frac{E}{T}\big)\label{eq:rhok}\end{eqnarray}

\begin{eqnarray}
 &  & E\frac{dR_{K^{*}+K\rightarrow\pi+\gamma}}{d^{3}p}=\qquad\qquad\label{eq:K*K}\\
 &  & \qquad T^{3.7}exp\big(\frac{-(6.096T^{1.889}+1.0299)}{(2TE)^{(-1.613T^{2.162}+0.975)}}-0.96\frac{E}{T}\big)\nonumber \end{eqnarray}
 Parameterisations for $K^{*}\rightarrow K+\pi+\gamma$ and $K+K\rightarrow\rho+\gamma$
do not appear because their rates have been found to be negligible.

Hadrons are composite objects, so they may need vertex form factors
to simulate finite hadronic size effect, in particular at high momentum
transfer. How much will form factor influence the results? Form factors
are a very delicate subject, especially when electromagnetism and
its gauge invariance are involved. For those nonstrange reaction channels
as originally studied by Kapusta \textit{et al.} \cite{Kapusta1991},
considering form factors provide a typical net suppression compared
to the bare graphs by an appreciable factor of $\sim3$ at photon
energies around $E\simeq2.5$~GeV. The reduction of the rate introduced
by Rapp \textit{et al.} \cite{Rapp2004} in $2\sim3$~GeV region
of photon energies amount to a factor of $\sim4$, confirming roughly
Kapusta's findings. %

\begin{figure}
\includegraphics[scale=0.75]{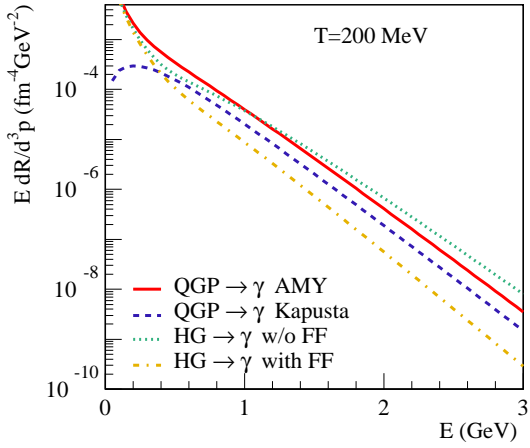}

\caption{\label{fig:Rate1} Photon emission rate from QGP and from HG. The
dashed line represents Kapusta's \textit{et al}\cite{Kapusta1991}
calculations of photon emission rates from a QGP at 200~MeV via $2\rightarrow 2$
processes, $\ie,$ QCD Compton process and quark-antiquark annihilation.
The solid line refers to AMY\cite{AMY2001}, where in addition the
LPM contribution was taken into account. The dotted line is a sum
of photon emission rates from a HG via different reactions (bare graphs)
shown in fig. \ref{fig:Rate2}. The dotted dashed line stands for
photon emission rates from a HG including hadronic form factors.}
\end{figure}

In ref. \cite{Arleo2004}, for each hadronic vertex appearing in
the amplitudes, it is proposed to take hadronic form factor for $t$-channel
meson $X$-exchange according to \begin{equation}
F(\overline{t})=\left(\frac{2\Lambda^{2}}{2\Lambda^{2}-\overline{t}}\right)^{2}\label{eq:fft}\end{equation}
 with $\Lambda$=1GeV and $\bar{t}$ being the average momentum transfer
$\overline{t}=-2Em_{X}$, where $E$ is the photon energy and $m_{X}$
the mass of the hadron $X$. In our calculation, we follow this procedure:
The photon emission rate is taken to be the sum of all terms in eqs.
(8-14), with each term multiplied by $F^{4}(\overline{t})$. The corresponding
curve is shown in Fig.\ref{fig:Rate1} as the dashed-dotted line.
The dotted line is the photon emission rate without considering hadronic
form factors. %%\begin{eqnarray}
%%E\frac{dR^{HG\rightarrow\gamma}}{d^{3}p} & = & F^{4}(\bar{t})E\frac{dR_{\pi+\rho\rightarrow\pi+\gamma}}{d^{3}p}\label{eq:rate_HG}\\
%% & + & F^{4}(\bar{t})E\frac{dR_{\pi+\pi\rightarrow\rho+\gamma}}{d^{3}p}\nonumber \\
%% & + & F^{4}(\bar{t})E\frac{dR_{\rho\rightarrow\pi+\pi+\gamma}}{d^{3}p}\nonumber \\
%% & + & F^{4}(\bar{t})E\frac{dR_{\pi+K^{*}\rightarrow K+\gamma}}{d^{3}p}\nonumber \\
%% & + & F^{4}(\bar{t})E\frac{dR_{\pi+\pi\rightarrow\rho+\gamma}}{d^{3}p}\nonumber \\
%% & + & F^{4}(\bar{t})E\frac{dR_{\rho+K\rightarrow K+\gamma}}{d^{3}p}\nonumber \\
%% & + & F^{4}(\bar{t})E\frac{dR_{K^{*}+K\rightarrow\pi+\gamma}}{d^{3}p}.\nonumber \end{eqnarray}
The form factors of Eq.(\ref{eq:fft}) makes the suppression stronger
compared to \cite{Rapp2004,Kapusta1991}. For example, at a photon
energy of $E=2.5$~GeV, the suppression factor is about $10$ for
$\pi$-exchange contributions, and even $625$ for $K$-exchange processes.
The suppression are even larger for higher photon energies. So the
hadronic emission rate is reduced considerably after including the
hadronic form factor as in Eq.(\ref{eq:fft}). The HG rate is therefore
much smaller than QGP one (full line).

\section{Thermal photons from the expanding hot and dense matter }

\begin{figure*}
\includegraphics[scale=0.75]{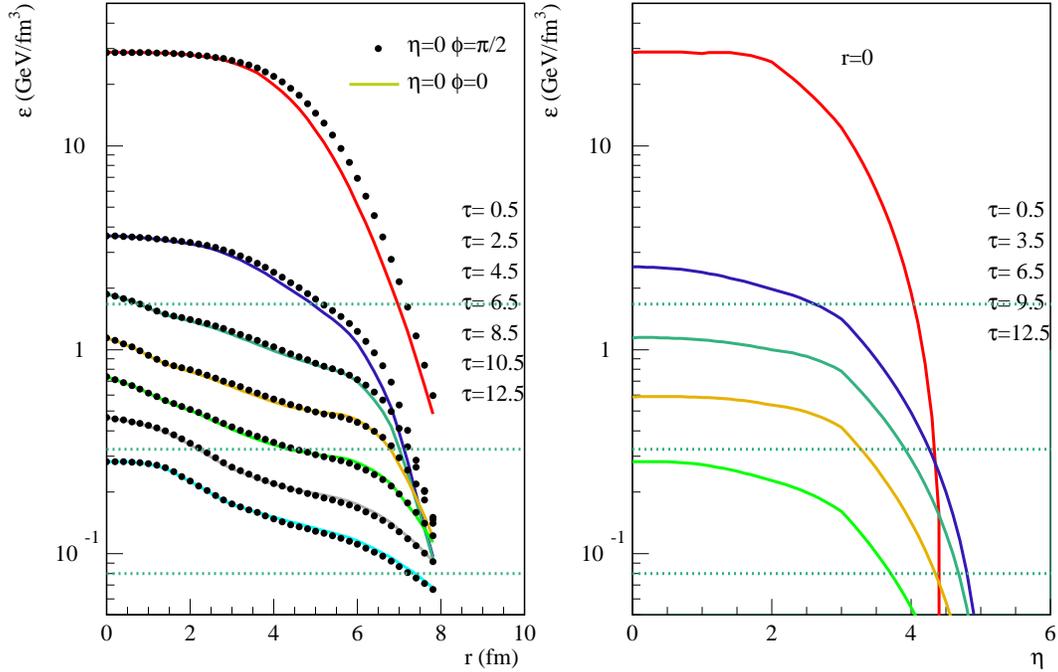}

\caption{\label{fig:energy-density} The energy density of the hot dense matter
created in the most 10\% central Au-Au collisions at 200~AGeV, is
plotted as function of the transverse space variable $r$. Curves
from up top to bottom correspond to different proper times, $\tau$=0.5,
2.5, ... (fm/c). The initial time is $\tau_{0}=0.5\mathrm{fm}/\mathrm{c}$.
The solid lines and the corresponding dotted lines refer to $\phi=0$
and $\phi=\pi/2$, respectively. The two upper horizontal dotted define
the mixed phase, the lower horizontal line the freeze-out energy density
$\varepsilon_{3}=0.08$~GeV/fm$^{3}$.}
\end{figure*}

The expanding thermalized matter is treated by employing three-dimensional
hydrodynamics. We compute initial conditions at some given proper
time $\tau_{0}$, expressed via energy density $\varepsilon(\tau_{0})$,
net flavor density $f_{q}(\tau_{0})$, and collective velocity $\vec{v}(\tau_{0})$,
by employing the EPOS model \cite{epos1,epos2}. The hydrodynamic
evolution is realized using SPheRIO \cite{SPHERIO}, which is a {}``Smoothed
Particle Hydrodynamics'' implementation, a method originally developed
in astrophysics, and later adapted to relativistic heavy ion collisions
. The three-dimensional hydrodynamics describes the space-time evolution
of the hot dense matter created in heavy ion collisions, via the 3-velocity
$\vec{v}$, the energy density $\varepsilon$, the entropy density
$s$ and the baryon number density $n_{B}$, as functions of the space-time
position $(\eta,\tau,r,\phi)$, with $\eta$ being the space-time
rapidity, and $r,\phi$ the transverse coordinates. In this paper,
we consider central AuAu collisions (10\% most central events) at
200~AGeV. The corresponding results of a hydrodynamical evolutions
is shown in Fig.\ref{fig:energy-density}, where we plot the $r$--dependence
of $\varepsilon$ at $\eta=0$ for different values of $\phi$ and
$\tau$. The initial time is $\tau_{0}=0.5\,\mathrm{fm}/\mathrm{c}$.
The solid lines and the corresponding dotted lines refer to $\phi=0$
and $\phi=\pi/2$, respectively. The three horizontal dotted lines
are the the energy densities $\varepsilon_{1}=1.675$~GeV/fm$^{3}$
and $\varepsilon_{2}=0.325$~GeV/fm$^{3}$ , limiting the mixed phase,
and the freeze-out energy density $\varepsilon_{3}=0.08$~GeV/fm$^{3}$..

\begin{figure}
\includegraphics[scale=0.75]{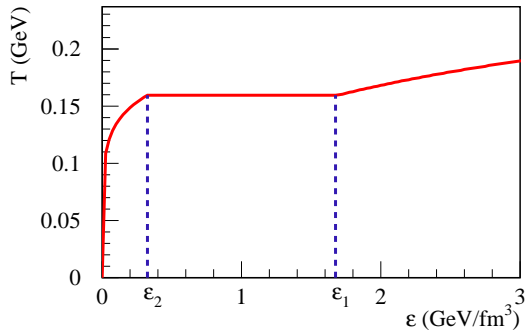}

\caption{\label{fig:eos} The relation between temperature and energy density.
A first-order phase transition happens when $\epsilon(x,y,\eta,\tau)\in(\epsilon_{2},\,\epsilon_{1})$
with space-time point $(x,y,\eta,\tau)$ in a mixed phase.}
\end{figure}

The relation between energy density and temperature, $T=T(\varepsilon)$,
as used in SPheRIO is shown in Fig.\ref{fig:eos} (and tabulated for
later use).

We define $f_{{\rm \textrm{QGP}}}$ to be the fraction of matter in
the QGP phase and $f_{{\rm \textrm{HG}}}$ as the corresponding fraction
in the HG phase, at each space-time point $(x,y,\eta,\tau)$. We have
obviously $f_{{\rm \textrm{QGP}}}=1$, $f_{{\rm \textrm{HG}}}=0$,
if the energy density is bigger than $\varepsilon_{1}$, $f_{{\rm \textrm{QGP}}}=0$,
$f_{{\rm \textrm{HG}}}=1$, if the energy density is between $\varepsilon_{3}$
and $\varepsilon_{2}$, and $f_{{\rm \textrm{QGP}}}=0$, $f_{{\rm \textrm{HG}}}=0$,
if the energy density is smaller than $\varepsilon_{3}$. In the mixed
phase ($\varepsilon_{2}<\varepsilon$$<\varepsilon_{1}$, we have
$s_{1}f_{{\rm \textrm{QGP}}}+s_{2}f_{{\rm \textrm{HG}}}=s$, with
$s_{1}$ and $s_{{\rm 2}}$ being the entropy densities corresponding
to $\varepsilon_{1}$ and $\varepsilon_{2}$, $s$ is the total entropy,
which is simply a linear function of $\varepsilon$ , like $s=a\epsilon+b$.
Then $f_{{\rm \textrm{QGP}}}$ and $f_{{\rm \textrm{HG}}}$ are linear
in $\varepsilon$ as well, namely\begin{equation}
f_{{\rm \textrm{QGP}}}=(1-f_{\mathrm{HG}})=\frac{\varepsilon-\varepsilon_{2}}{\varepsilon_{1}-\varepsilon_{2}}.\,\label{eq:fQGP}\end{equation}

\begin{figure}
\includegraphics[scale=0.75]{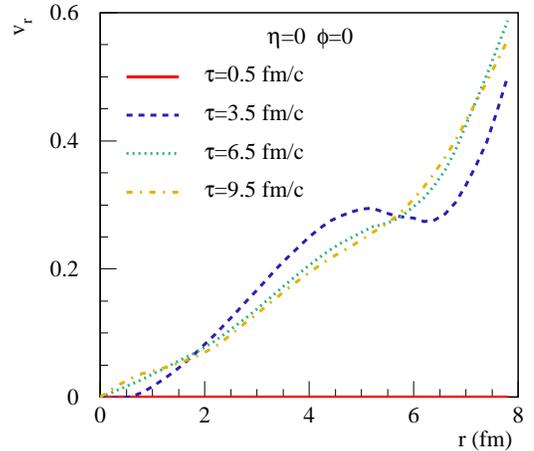}

\caption{\label{fig:ig1_vr} Transverse velocity $v_{r}$ of the hydrodynamic
expansion for the hot dense matter created in the most 10\% central
Au-Au collisions at 200~AGeV.}
\end{figure}

The photon emission rates discussed in the previous chapter are obtained
in the local rest frame, so we should better write \begin{equation}
\Gamma(E^{*},T)=E^{*}\frac{dR}{d^{3}p^{*}}(E^{*},T)=E^{*}\frac{dN}{d^{3}p^{*}d^{4}x^{*}},\label{eq:ratedef}\end{equation}
 where quantities viewed in the local rest frame are decorated with
the superscript $*$. Bare symbols (without {*}) refer to the laboratory
frame. We need the photon spectrum observed in the laboratory, which
is given as\begin{eqnarray}
\frac{dN}{dyd^{2}\pt} & = & E\frac{dN}{d^{3}p}=\int d^{4}x\Gamma(E,T)\nonumber \\
 & = & \int d^{4}x\Gamma(E^{*},T)\nonumber \\
 & = & \int d\tau dxdyd\eta\tau\Gamma(E^{*},T),\label{eq:E*a}\end{eqnarray}
 with\begin{equation}
\Gamma(E^{*},T)=f_{{\rm QGP}}\Gamma^{{\rm QGP}\rightarrow\gamma}(E^{*},T)+f_{{\rm HG}}\Gamma^{{\rm HG}\rightarrow\gamma}(E^{*},T)\label{eq:rate2parts}\end{equation}
 where $\Gamma^{{\rm QGP}\rightarrow\gamma}(E^{*},T)$ and $\Gamma^{{\rm HG}\rightarrow\gamma}(E^{*},T)$
are the photon emission rates from QGP phase and from HG phase as
discussed in the previous chapter. The center-of-mass energy $E^{*}$
in eq.(\ref{eq:E*a}) is related to the photon momenta in the observer
frame (appearing on the l.h.s. of Eq.(\ref{eq:E*a})) as \begin{equation}
E^{*}=\gamma E-\gamma\vec{v}(\tau,x,y,\eta) \cdot \vec{p},\end{equation}
 with\begin{equation}
\gamma=\frac{1}{\sqrt{1-\left|\vec{v}(\tau,x,y,\eta)\right|^{2}}} ,\end{equation}
 where $\vec{v}(\tau,x,y,\eta)$ is the flow velocity at a given space-time
point. The flow rapidity $y=\frac{1}{2}\ln(1+v_{z})/(1-v_{z})$ is
roughly equal to the space-time rapidity (as in the Bjorken model).
The radial dependence of the transverse velocity $v_{r}$ is shown
in Fig.\ref{fig:ig1_vr}, for central Au-Au collisions at 200~AGeV.
The large transverse flows at large radii will boost thermal photons
to higher $\pt$ regions.

\begin{figure}
\includegraphics[scale=0.75]{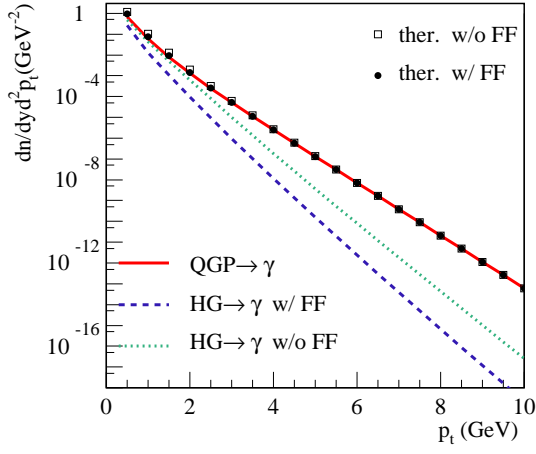}

\caption{\label{fig:ig1_th} Thermal photons from the most 10\% central Au-Au
collisions at 200~AGeV with initial time $\tau_{0}$=0.5~fm/c. The
contributions from the two phases are presented separately: QGP phase
(solid line) and HG phase (dashed line). The upper limit of the contribution
from HG phase, $\ie$, without consideration of hadron form factor,
is presented as dotted line. The total thermal contribution
is plotted in the two cases, with hadron form factor (full circles
) and without hadron form factor (empty squares).}
\end{figure}

In Fig.\ref{fig:ig1_th} we present the transverse momentum spectra
of thermal photons produced in the 10\% most central Au-Au collisions
at 200~AGeV with an initial time of $\tau_{0}$=0.5~fm/c. The contributions
from the two phases are presented separately: QGP phase (solid line)
and HG phase (dashed line). We can see thermal contribution is dominated
by photons from QGP phase, which is much bigger than the upper limit
of the contribution from HG phase, $\ie$, without consideration of
hadron form factors (dotted line). The reason is simply that the QGP
phase is \char`\"{}hotter\char`\"{} compared to the HG phase. Therefore,
hadron form factors makes very little difference concerning the total
thermal contribution, as seen in Fig.\ref{fig:ig1_th}, when comparing
the complete results with form factor (full circles) and without form
factor (empty squares).

\section{Hard photons from primordial N-N scattering, including shadowing
and isospin mixing}

\begin{figure}
\includegraphics[scale=0.75]{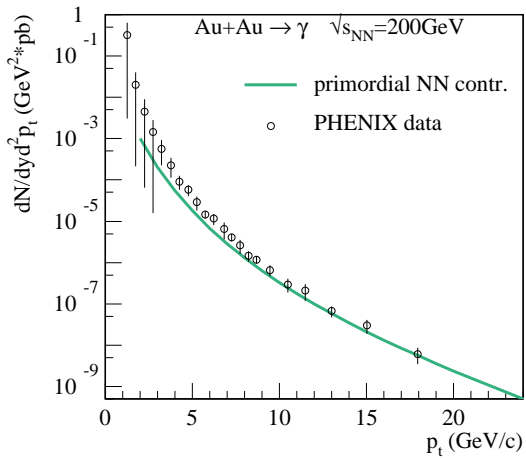}

\caption{\label{fig:fig} The contribution of primordial NN collisions to
direct photon production in the 10\% most central Au-Au collisions
at 200~AGeV. Isospin mixing and nuclear shadowing are considered.
Empty circles present PHENIX data\cite{PHENIX data}.}
\end{figure}

Here we present the leading order perturbative calculation of hard
photon production, namely the hard photons from primordial N-N scattering.
Higher order contributions are related to the production and propagation
of jets, so we treat them later. The spectrum of hard photons from
a collision between nuclei $A$ and nuclei $B$ is

\begin{eqnarray}
 &  & \frac{dN^{AB\rightarrow\gamma}}{dyd^{2}\pt}=\sum_{{\displaystyle ab}}\int dx_{a}dx_{b}G_{a/A}(x_{a},M^{2})G_{b/B}(x_{b},M^{2})\nonumber \\
 &  & \qquad\qquad\frac{\hat{s}}{\pi}\frac{d\sigma}{d\hat{t}}(ab\rightarrow\gamma+X)\delta(\hat{s}+\hat{t}+\hat{u})\label{eq:ABtogamma}\end{eqnarray}
 where $G_{a/A}(x_{a},M^{2})$ and $G_{b/B}(x_{b},M^{2})$ are parton
distribution functions for nuclei $A$ and $B$. We take MRST 2001
LO parton distributions for protons\cite{MRST0201}. Nuclear shadowing
effects are taken into account by using EKS98 scale dependent nuclear
ratios $R_{a}^{{\rm \textrm{EKS}}}(x,A)$\cite{EKS98}. The mixed
isospin in nuclei with mass $A$, neutron number $N$ and proton number
$Z$ is taken into account as\begin{equation}
G_{a/A}(x)=A[\frac{N}{A}G_{a/N}(x)+\frac{Z}{A}G_{a/P}(x)]R_{a}^{{\rm \textrm{EKS}}}(x,A).\label{eq:PdisA}\end{equation}
 The elementary cross sections after color sum and spin average are
given as \cite{Owens1987}

\begin{equation}
\frac{d\sigma}{d\hat{t}}(qg\rightarrow q\gamma)=-\frac{1}{3}\frac{\pi\alpha\alpha_{s}}{\hat{s}^{2}}e_{q}^{2}(\frac{\hat{u}}{\hat{s}}+\frac{\hat{s}}{\hat{u}})\label{eq:qgtogamma}\end{equation}
 \begin{equation}
\frac{d\sigma}{d\hat{t}}(q\bar{q}\rightarrow g\gamma)=\frac{8}{9}\frac{\pi\alpha\alpha_{s}}{\hat{s}^{2}}e_{q}^{2}(\frac{\hat{u}}{\hat{t}}+\frac{\hat{t}}{\hat{u}}),\label{eq:qqbartogamma}\end{equation}
 with \begin{equation}
\alpha_{s}=\frac{12\pi}{33-2N_{f}}\frac{1}{\ln(Q^{2}/\Lambda_{{\rm QCD}}^{2})},\end{equation}
 and $\Lambda_{{\rm QCD}}=200$~MeV. The 4-momenta of the incoming
particles ($p_{a}$, $p_{b}$) and of the photon ($p_{\gamma}$) in
the center-of-mass are

\begin{equation}
p_{a}=(x_{a}\frac{\sqrt{s}}{2},\,0,\,0,\, x_{a}\frac{\sqrt{s}}{2}),\; p_{b}=(x_{b}\frac{\sqrt{s}}{2},\,0,\,0,\,-x_{b}\frac{\sqrt{s}}{2})\label{eq:a1}\end{equation}
 and\begin{equation}
p_{\gamma}=(\pt\cosh y,\vec{\pt},\pt\sinh y).\label{eq:gamma1}\end{equation}
 The Mandelstam variables are then given as

\begin{equation}
\hat{s}=(p_{a}+p_{b})^{2}=x_{a}x_{b}s\label{eq:shat}\end{equation}
 \begin{equation}
\hat{t}=(p_{a}-p_{\gamma})^{2}=-x_{a}\sqrt{s}\pt\exp(-y)\label{eq:that}\end{equation}
 \begin{equation}
\hat{u}=(p_{b}-p_{\gamma})^{2}=-x_{b}\sqrt{s}\pt\exp(y).\label{eq:uhat}\end{equation}
 We set the factorization scale $M$ and renormalization scale $Q$
to be $\pt$. Due to the $\delta$-function in Eq.(\ref{eq:ABtogamma}),
the $x_{b}$-integration is trivial, leading to \begin{equation}
x_{b}=\frac{x_{a}x_{\bot}\exp(-y)}{2x_{a}-x_{\bot}\exp(y)},\end{equation}
 with $x_{\bot}=2\pt/\sqrt{s}$. The condition $x_{b}<1$ requires
$x_{a}\in(x_{{\rm min}},1)$ with $x_{{\rm min}}=x_{\bot}\exp(y)/(2-x_{\bot}\exp(-y))$.
So

\begin{eqnarray}
 &  & \frac{dN^{AB\rightarrow\gamma}}{dyd^{2}\pt}=\sum_{{\displaystyle ab}}\int_{x_{{\rm min}}}^{1}dx_{a}G_{a/A}(x_{a},M^{2})G_{b/B}(x_{b},M^{2})\nonumber \\
 &  & \qquad\qquad\frac{1}{x_{a}s-\sqrt{s}\pt\exp(y)}\frac{\hat{s}}{\pi}\frac{d\sigma}{d\hat{t}}(ab\to\gamma+X).\label{eq:LOphoton}\end{eqnarray}
 In Fig.\ref{fig:fig}, we plot the corresponding spectrum, for central
Au-Au collisions at 200~AGeV. Empty circles present PHENIX data\cite{PHENIX data}.
At high $\pt$ region, the contribution of primordial NN collisions
is very close the experimental data. %

\section{Jet photon conversion, including jet energy loss}

When jets propagate in the hot dense matter created in heavy ion collisions,
they also interact with the matter and produce direct photons via the
Compton process $qg\rightarrow q\gamma$ and the quark-antiquark annihilation
process $q\bar{q}\rightarrow g\gamma$.
We denote the momenta of the jet, the thermal parton, and the
photon by $p_{{\rm jet}}$, $p_{{\rm th}}$ and $p$, respectively.
The leading order QCD Compton and annihilation cross sections are
peaked in the forward and backward directions. In the laboratory frame
we have $|p_{{\rm jet}}|\gg|p_{{\rm th}}|\sim T$, where $T$ is the
temperature of the plasma. For high energy photons, i.e. $|p|\gg T\ $,
this implies that $|p|\approx|p_{{\rm jet}}|$. That is why we call
this process a conversion of a jet into a photon with similar momentum.
The rate of photon production by annihilation and Compton scattering
of jets in the medium can be approximated as \cite{WangCY}\cite{Fries2005}\begin{equation}
E\frac{dN}{d^{3}pd^{4}x}=\frac{\alpha\alpha_{s}}{4\pi^{2}}\sum_{q}e_{q}^{2}f_{q}(p,x)T^{2}\left[\ln\frac{4E_{\gamma}T}{m_{{\rm th}}^{2}}+C\right],\label{eq:jpc_rate}\end{equation}
 with $C=-1.916$, $m_{{\rm th}}^{2}=g^{2}T^{2}/6$, and where $\alpha_{s}=g^{2}/(4\pi)$
and $\alpha$ are the strong and the electromagnetic couplings. Th
subscript {}``$q$'' denotes all light quark and antiquark species
with charge $e_{q}$, and $f_{q}=f_{q}(\vec{p},x)$ is the phase-space
density of partons of flavor $q$. It is worth emphasizing that the conversion
property of the process is reflected in eq.(\ref{eq:jpc_rate}) by
the fact that the photon spectrum is directly proportional to the
parton spectrum $f_{q}$. Then at midrapidity the jet-photon conversion
contribution to the direct photon production in the most 10\% Au-Au
collisions is gained via the integration over the space-time evolution
of the hot dense matter in QGP phase:\begin{eqnarray}
 &  & \frac{dN^{{\rm {\rm jet+QGP}}\rightarrow\gamma}}{dyd^{2}\pt}\label{eq:jpc_int}\\
 &  & \quad=\int E\frac{dN}{d^{3}pd^{4}x}f_{{\rm QGP}}(x,y,\eta,\tau)dxdyd\eta\tau d\tau.\nonumber \end{eqnarray}
 To get $f_{q}$, the phase space densities of quarks and antiquarks,
we have to fix the geometry of jet formation. The jet production from
primordial N-N scattering is assumed to happen at the same proper
time $\tau=0$. Then at $\tau=0$, the phase space distribution of
partons of type $q$ is \begin{equation}
f_{q_{i}}(\vec{p},\vec{r},\tau=0)\propto T_{A}(x-\frac{b}{2},y)T_{B}(x+\frac{b}{2},y)\delta(z),\label{eq:space dis}\end{equation}
 where $b$ in the impact parameter and $T_{A}$ and $T_{B}$ are
thickness functions of nuclei $A$ and $B$. The $\delta$-function
is motivated by the strong Lorentz contraction of the colliding nuclei.
To have a simple form for a central $A$-$A$ collision, i.e. most
10\% central Au-Au collision, we take the approximation

\begin{eqnarray}
 &  & f_{q}(\vec{p},\vec{r},\tau=0)=\frac{(2\pi)^{3}}{E}\frac{dN^{AB\rightarrow jet(q)}}{dyd^{2}\pt}\label{eq:fq1}\\
 &  & \qquad\qquad\frac{2}{\pi R_{\bot}^{2}}(1-\frac{r_{\bot}^{2}}{R_{\bot}^{2}})\theta(R_{\bot}-r_{\bot})\delta(z)\nonumber \end{eqnarray}
 where $R_{\bot}$is the radius of nuclei Au, and with $r_{\bot}=\sqrt{x^{2}+y^{2}}$.
The momentum distribution is calculated as \begin{eqnarray}
 &  & \frac{dN^{AB\rightarrow{\rm jet}}}{dyd^{2}\pt}\label{eq:LOjet}\\
 &  & \quad=\sum_{{\displaystyle abcd}}\int_{x_{{\rm {\rm min}}}}^{1}dx_{a}K_{{\rm jet}}G_{a/A}(x_{a},M^{2})G_{b/B}(x_{b},M^{2})\nonumber \\
 &  & \qquad\qquad\qquad\frac{1}{x_{a}s-\sqrt{s}\pt\exp(y)}\frac{\hat{s}}{\pi}\frac{d\sigma}{d\hat{t}}(ab\rightarrow cd),\nonumber \end{eqnarray}
 which is very similar to eq.(\ref{eq:LOphoton}). Here, the cross
sections of all possible partonic processes $ab\rightarrow cd$ make
totally 127 terms\cite{Owens1987}. $K_{jet}$=2 is used to take
into account higher order contributions to jet production in our calculation%
\footnote{$K_{jet}$ can be extracted from the ratio of experimental jet date
to leading order calculation. We find $K_{jet}\approx$2 via pp (p$\bar{p}$)
collisions at energy range from 27.4GeV to 630GeV. %
}. The phase space distribution of jet satisfies the normalization
\begin{equation}
\int f_{q}(\vec{p},\vec{r},\tau=0)\frac{d^{3}rd^{3}p}{(2\pi)^{3}}=N_{q},\label{eq:fqi}\end{equation}
 where $N_{q}$ is the number of jets of type $q$.

If we ignore the jet energy loss due to the interaction between jets
and matter, then at any $\tau>0$, the phase space distribution of
jets is\begin{eqnarray}
f(\vec{p},\vec{r},\tau) & = & \int d^{3}r_{0}f(\vec{p},\vec{r}_{0},\tau=0)\delta(\vec{r}-\vec{r}_{0}-\vec{v}t)\nonumber \\
 & = & f(\vec{p},\vec{r}-\vec{v}t,0)\label{eq:fq2}\end{eqnarray}
 where $\vec{v}$ is the velocity of a jet, $\vec{v}=\vec{p}/E$,
and $t=\tau cosh(\eta)$ with $\eta$ being the space time rapidity.

If we consider the modification of the jet energy and its momentum
due to the interaction between jets and hot dense matter, then the
phase space distribution $f(\vec{p},\vec{r},\tau)$ of jets should
be replaced by

\begin{eqnarray}
 &  & \int d^{3}r_{0}d^{3}p_{0}f(\vec{p}_{0},\vec{r}_{0},0)\delta(\vec{p_{0}}-\frac{\vec{p}}{E}\Delta E-\vec{p})\delta(\vec{r}-\vec{r}_{0}-\vec{v}t)\nonumber \\
 &  & =\int d^{3}p_{0}f(\vec{p}_{0},\vec{r}-\vec{v}t,0)\delta(\vec{p_{0}}-\frac{\vec{p}}{E}\Delta E-\vec{p})\label{eq:fq3}\end{eqnarray}
 with $E=|\vec{p}|$, and where $\Delta E$ is the energy loss of
a jet propagates from the formation point $(\vec{r}_{0},0)$ to the
jet-medium interaction point $(\vec{r},\tau)$. In static matter,
one has \cite{Baier97}\begin{equation}
\Delta E_{c}=\frac{L}{\lambda_{c}}\epsilon_{c},\label{eq:Eloss1}\end{equation}
 with $\epsilon_{c}=\alpha_{s}\sqrt{\mu^{2}E/\lambda}_{c}$, and where
the mean free path $\lambda_{c}$ of the jet in the medium, and $\mu=gT$,
are all temperature dependent quantities. The index {}``$c$'' refers
to the jet type (quark or gluon). Due to the space-time evolution
of temperature as $T=T(x,y,\eta,\tau)$, we have to replace $L$ in
Eq.(\ref{eq:Eloss1}) by the integration $\int_{0}^{t}v\, f_{\mathrm{QGP}}dt'$
along the jet's trajectory in the QGP plasma. Light-flavour quarks and gluons
are massless, with $v=1$, and so we get \begin{eqnarray}
\Delta E & = & \int_{0}^{t}\frac{\epsilon_{c}(T(x,y,\eta,\tau))}{\lambda_{c}(T(x,y,\eta,\tau))}f_{\mathrm{QGP}}(x,y,\eta,\tau)dt'.\label{eq:Eloss2}\end{eqnarray}
 The mean free path $\lambda_{c}$ is given as \begin{equation}
\lambda_{g}^{-1}=\sigma_{gq}\rho_{q}+\sigma_{gg}\rho_{g},\label{eq:labdag}\end{equation}
 \begin{equation}
\lambda_{q}^{-1}=\sigma_{qq}\rho_{q}+\sigma_{qg}\rho_{g},\label{eq:lambdaq}\end{equation}
 where the cross sections are given as \cite{GW94} $\sigma_{i}=C_{i}\frac{\alpha\pi}{T^{2}}$,
with $C_{qq}=\frac{4}{9}$, $C_{qg}=1$ and $C_{gg}=\frac{9}{4}$,
and where $\rho_{q}$ and $\rho_{g}$ are the thermal parton densities.

\begin{figure}
\includegraphics[scale=0.75]{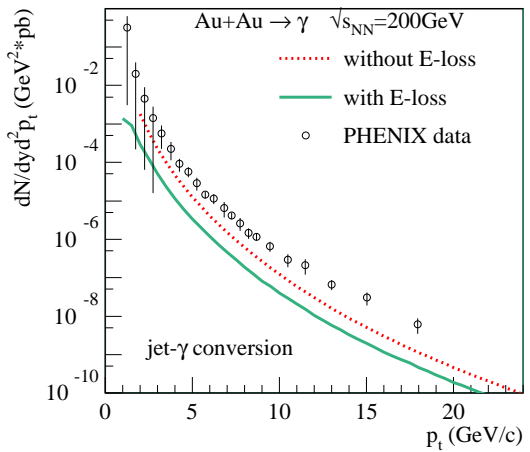}

\caption{\label{fig:fig2} Jet-photon conversion contribution to direct photon
production in the 10\% most central Au-Au collisions at 200~AGeV.
Dotted line: jets traverse the hot dense matter without energy loss;
solid line: jets traverse the hot dense matter with energy loss; Empty
circles present PHENIX data\cite{PHENIX data}. The effect from different
hydrodynamic initial times $\tau_{0}$ is also considered. However,
results with $\tau_{0}$=0.5~fm/c and 1~fm/c are indistinguishable. }
\end{figure}

Fig.\ref{fig:fig2} shows the jet-photon conversion contribution to
direct photon production in central Au-Au collisions at 200~AGeV.
We show the results with and without energy loss, together with PHENIX
data\cite{PHENIX data}. We can see that considering energy loss
can indeed suppress the jet-photon conversion contribution, by about
a factor of three. Obviously, higher jet energies are needed to produce
a photon with a given energy, in case of jet energy loss in the medium.
The effect from different hydrodynamic initial times $\tau_{0}$ is
also considered. However, results with $\tau_{0}$=0.5~fm/c and 1~fm/c
are indistinguishable. Our results also agree well with the earlier
work by Turbide\cite{Turbide2005}, although they use one dimensional
hydrodynamics.

\section{Bremsstrahlung photons, including jet energy loss}

We discussed earlier leading order photon production from elementary
NN collisions. However, also higher order diagrams give important
contributions. Here, we discuss {}``bremsstrahlung photons'' from
jets, also referred to as {}``jet fragmentation''. We can treat
the bremsstrahlung contribution via parton fragmentation functions
$D_{\gamma/c}(z,Q^{2})$ being the probability for obtaining a photon
from a jet $c$, where the photon carries a fraction $z$ of the jet's
momentum. The effective fragmentation functions for obtaining photons
from partons can be calculated. The leading order result is

\begin{equation}
zD_{\gamma/q}(z,Q^{2})=e_{q}^{2}\frac{\alpha}{2\pi}[1+(1-z)^{2}]\ln(Q^{2}/\Lambda^{2})\label{eq:q-gammaLO}\end{equation}
 and\begin{equation}
zD_{\gamma/g}(z,Q^{2})=0,\label{eq:g-gammaLO}\end{equation}
 where $e_{q}$ is the fractional charge of the quark $q$. The photon
fragmentation functions evolve with $Q^{2}$ just as the usual hadronic
fragmentation functions do, as a result of gluon bremsstrahlung and
$q\bar{q}$ pair production. The resulting evolution equations are

\begin{eqnarray}
 &  & \frac{dD_{\gamma/q_{i}}(z,Q^{2})}{dt}\label{eq:q-frag}\\
 &  & \qquad=\frac{\alpha_{s}(Q^{2})}{2\pi}\int_{z}^{1}\frac{dy}{y}[D_{\gamma/q_{i}}(y,Q^{2})P_{qq}(z/y)\nonumber \\
 &  & \qquad\qquad+D_{\gamma/g}(z,Q^{2})P_{gq}(z/y)],\nonumber \end{eqnarray}
 and

\begin{eqnarray}
 &  & \frac{dD_{\gamma/g}(z,Q^{2})}{dt}\label{eq:g-frag}\\
 &  & \qquad=\frac{\alpha_{s}(Q^{2})}{2\pi}\int_{z}^{1}\frac{dy}{y}[\sum_{i=1}^{2N_{f}}D_{\gamma/q_{i}}(y,Q^{2})P_{qg}(z/y)\nonumber \\
 &  & \qquad\qquad+D_{\gamma/g}(z,Q^{2})P_{gg}(z/y)]\nonumber \end{eqnarray}
where $P_{qq}$, $P_{gq}$,$P_{qg}$ and $P_{qq}$ are splitting functions appearing in the DGLAP equations. The parameterized solutions from Owens\cite{Owens1987} are

\begin{eqnarray}
 &  & zD_{\gamma/q_{i}}(z,Q^{2})\label{eq:q-gamma0}\\
 &  & \qquad=\frac{\alpha}{2\pi}[e_{i}^{2}\frac{2.21-1.28z-1.29z^{2}}{1-1.63\ln(1-z)}\nonumber \\
 &  & \quad\qquad+0.002(1-z)^{2}z^{-1.54}]\ln(Q^{2}/\Lambda^{2}),\nonumber \end{eqnarray}
 and\begin{equation}
zD_{\gamma/g}(z,Q^{2})=\frac{\alpha}{2\pi}0.0243(1-z)z^{-0.97}\ln(Q^{2}/\Lambda^{2}).\label{eq:g-gamma0}\end{equation}
 The bremsstrahlung contribution to direct photon production is then

\begin{equation}
\frac{dN^{jet\rightarrow\gamma}}{dyd^{2}\pt}=\sum_{c=g,q_{i}}\int dz_{c}\frac{dN^{AB\rightarrow jet_{c}}}{dyd^{2}\pt^{c}}\frac{1}{z_{c}^{2}}D_{\gamma/c}(z_{c},Q^{2}),\label{eq:jet-gamma0}\end{equation}
 with $\pt^{c}=\pt/z_{c}$ being the momentum carried by the jet $c$
before fragmentation, and $d^{3}p/E=z_{c}^{2}d^{3}p^{c}/E^{c}$. The
jet cross section can be obtained from Eq.(\ref{eq:LOjet}).

To take into account the energy loss of the jet in the QGP phase,
the bremsstrahlung contribution to direct photon production is modified.
We may use a modified fragmentation function\cite{XNWANG04}, given
as \begin{eqnarray}
 &  & D_{\gamma/c}(z_{c},Q^{2},\Delta E_{c})\label{eq:jet-gamma}\\
 &  & \qquad=(1-e^{-\frac{L}{\lambda_{c}}})[\frac{z_{c}'}{z_{c}}D_{\gamma/c}^{0}(z_{c}',Q^{2})+\frac{L}{\lambda_{c}}\frac{z_{g}'}{z_{c}}D_{\gamma/g}^{0}(z_{g}',Q^{2})]\nonumber \\
 &  & \qquad\qquad\qquad+e^{-\frac{L}{\lambda_{c}}}\: D_{\gamma/c}^{0}(z_{c},Q^{2}),\nonumber \end{eqnarray}
 where $z_{c}'=\pt/(\pt^{c}-\Delta E_{c})$ and $z_{g}'=(L/\lambda_{c})\,\pt/\Delta E_{c}$
are the rescaled momentum fractions carried by jet $c$ and the emitted
gluons before fragmentation, and where $D_{\gamma/q}^{0}$ and $D_{\gamma/g}^{0}$
are the original fragmentation functions, given in Eqs.(\ref{eq:q-gamma0})
and (\ref{eq:g-gamma0}). So a parton $c$ has the probability $\exp(-L/\lambda_{c})$
to fragment directly without interacting with the medium, and the
probability $1-\exp(-L/\lambda_{c})$ to interact with the medium
before fragmentation. The ratio $L/\lambda_{c}$ represents the number
of scatterings, and a gluon of energy $\epsilon_{c}$ is emitted each
time when a parton scatters with the medium.

As discussed earlier, the energy loss in the medium with a constant
temperature is $\Delta E_{c}=\frac{L}{\lambda_{c}}\epsilon_{c}$,
where $\lambda_{c}$ is the mean free path of jet $c$ in the medium
and $\epsilon_{c}=\alpha_{s}\sqrt{\mu^{2}E/\lambda_{c}}$. Those quantities
are temperature dependent. In our case, the temperature evolves with
space and time. So similar to Eq.(\ref{eq:Eloss2}), we replace $L/\lambda_{c}$
and $\Delta E_{c}$ by the corresponding mean values, namely \begin{equation}
\int_{\tau_{1}}^{\infty}\frac{1}{\lambda_{c}(T(x,y,\eta,\tau))}f_{QGP}(x,y,\eta,\tau)dt,\label{eq:sctt}\end{equation}
 \begin{equation}
\int_{0}^{\infty}\frac{\epsilon_{c}(T(x,y,\eta,\tau))}{\lambda_{c}(T(x,y,\eta,\tau))}f_{QGP}(x,y,\eta,\tau)dt.\label{eq:dEc}\end{equation}
 We take the mean energy loss per scattering $\lambda\Delta E_{c}/L$
as the energy carried by each emitted gluon.

\begin{figure}
\includegraphics[scale=0.75]{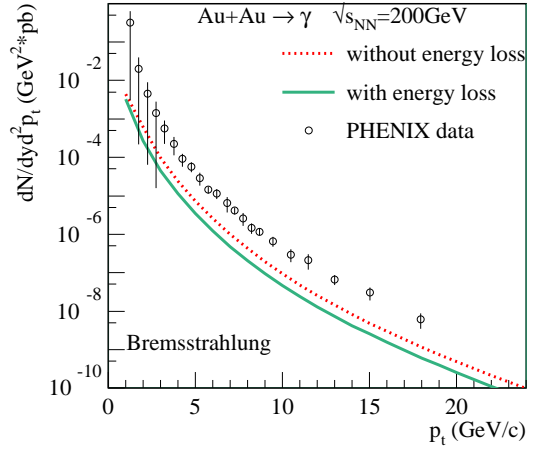}

\caption{\label{fig:fig3} Bremsstrahlungs (= jet fragmentation ) contribution
to direct photon production in the 10\% most central Au-Au collisions
at 200~AGeV. Dotted line: jets traverse the hot dense matter directly
without energy loss. Solid line: jets traverse the hot dense matter
with energy loss. Empty circles: PHENIX data\cite{PHENIX data}.}
\end{figure}

In Fig.\ref{fig:fig3}, we present the bremsstrahlung contribution
to direct photon production, with and without considering energy loss.
We can see that the photon production is considerably suppressed due
to the energy loss of the jets in the hot dense matter. The results
from two different initial times $\tau_{0}$=0.5~fm/c and 1~fm/c
is indistinguishable.

\section{results and discussion}

\begin{figure}
\includegraphics[scale=0.75]{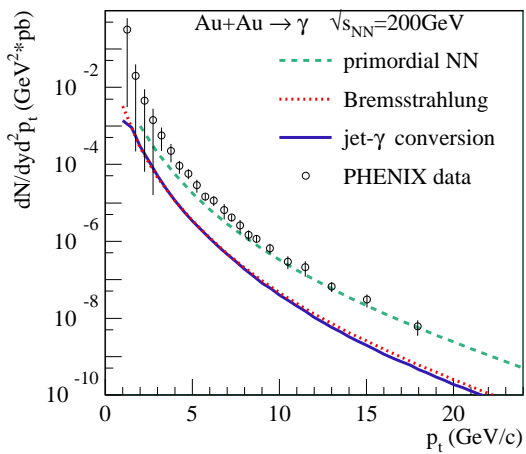}

\caption{\label{fig:fig4} Comparison of the different non-thermal contributions
photon production in the 10\% most central Au-Au collisions at 200~AGeV.
Dashed line: contribution from primordial NN collisions. Dotted line:
contribution from Bremsstrahlung. Solid line: jet-photon conversion
contribution. Jets energy loss in the hot dense matter is considered
both for bremsstrahlung photons and jet-photon conversion. Empty circles:
PHENIX data\cite{PHENIX data}.}
\end{figure}

In the following we are going to collect and discuss our results.
In Fig.\ref{fig:fig4} we compare the different non-thermal contributions
to photon production in central Au-Au collisions at 200~AGeV. We
show the contribution from primordial NN collisions (dashed line),
bremsstrahlung (or fragmentation) photons (dotted line), and photons
from jet-photon conversion (solid line). We have taken into account
energy loss of jets in the hot and dense matter, both for bremsstrahlung
photons and jet-photon conversion. We also plot PHENIX data\cite{PHENIX data},
as a reference. Jet-photon conversion contributes the same magnitude
as bremsstrahlung, but both are small compared to the hard photons
from primordial NN collisions.%

\begin{figure}
\includegraphics[scale=0.75]{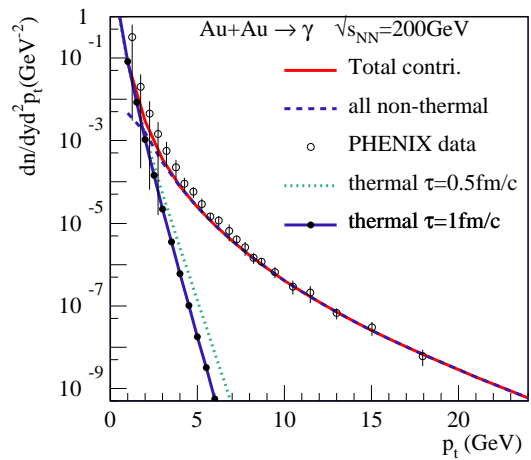}

\caption{\label{fig:fig5} The total contribution (solid line) is compared
with PHENIX data\cite{PHENIX data} (empty circles). Thermal contributions
from a hydrodynamic calculation with initial time $\tau_{0}$=0.5~fm/c
is plotted as dotted line, while $\tau_{0}$=1~fm/c is presented
as solid line with full circles. Different values of $\tau_{0}$ are
indistinguishable concerning the total contribution.}
\end{figure}

In Fig.\ref{fig:fig5}, the total contribution (solid line), including
all non-thermal and thermal contributions, is compared with PHENIX
data{[}PHENIX data] (empty circles). The thermal contribution from
a hydrodynamic calculation with an initial time $\tau_{0}$=0.5~fm/c
is plotted (dotted line) as well as the results of a calculation with
$\tau_{0}$=1~fm/c (solid line with full circles). Smaller $\tau_{0}$
gives a somewhat broader distribution, but this effect is invisible
in the total contribution. At very low $\pt$ , thermal production
dominates the direct photon production. However, the thermal spectra
decrease very fast with $\pt$, and become negligible for $\pt$$\geq$
4GeV/c. The hydrodynamic configuration also affects the jet-photon
conversion and the bremsstrahlung contribution. But this effect is
very weak: taking different initial times within the range {[}0.5,1]fm/c
make very little difference to the total contribution (invisible in
Fig.16). The complete contribution, containing all the thermal and
non-thermal contributions, follow quite well the PHENIX data.%

It is useful to compare the AuAu results with proton-proton, and here
it is convenient to study the nuclear modification factor, defined
to be ratio of the nuclear spectrum to the proton-proton one, divided
by the number $N_{{\rm coll}}$ of binary collisions. We use $N_{{\rm coll}}$=880
for the 10\% most central Au-Au collisions at 200~AGeV, treated in
this paper. We compute the proton-proton differential cross section
by using $\sigma_{{\rm inelastic}}^{pp}$ =40.83mb, and $K_{{\rm jet}}$=2,
and the unmodified photon fragmentation functions given in eq.(\ref{eq:q-gamma0},
\ref{eq:g-gamma0}). %
\begin{figure}
\includegraphics[scale=0.75]{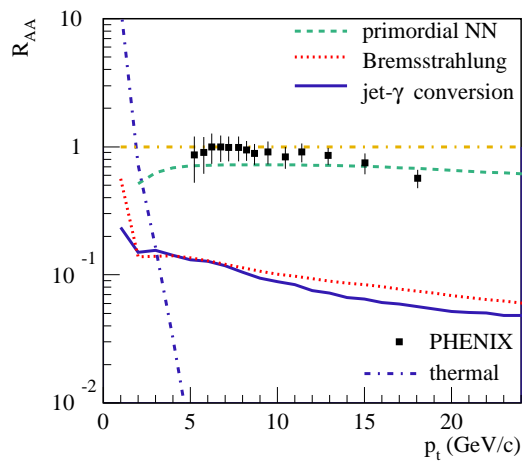}

\caption{\label{fig:fig6} Different non-thermal contributions and thermal
contribution to nuclear modification factor $R_{AA}$ is compared
with PHENIX data\cite{PHENIX data} (full squares). Dashed line:
the primordial NN contribution; Dotted line: bremsstrahlung; Solid
line: jet-gamma conversion; Dotted dashed line: thermal contribution. }
\end{figure}

In Fig.\ref{fig:fig6}, we plot the different non-thermal contributions,
namely the primordial NN contribution (dashed line), bremsstrahlung
(dotted line), and jet-gamma conversion (solid line). The contributions
from bremsstrahlung and jet-photon conversion are of the same magnitude,
but one order smaller than that the primordial NN curve. We also show
the rapidly falling thermal contribution(dotted dashed line). Above
5 GeV, only photons from primordial NN scattering contribute significantly.
This is why we look more closely to this latter contribution. %
\begin{figure}
\includegraphics[scale=0.75]{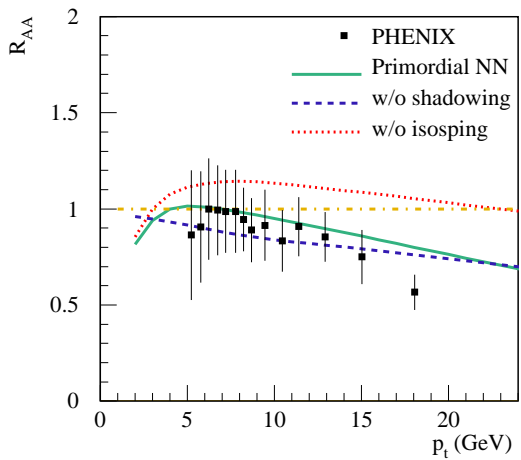}

\caption{\label{fig:fig7} A close look at primordial NN scattering contribute
to the nuclear modification factor $R_{AA}$. Solid line: the complete
primordial NN contributions, including isospin mixing and shadowing;
Dotted line: omitting isospin mixing (but considering shadowing);
Dashed line: omitting shadowing (but considering isospin mixing).
Full squares: the same data as in Fig.\ref{fig:fig6}.}
\end{figure}

In Fig.\ref{fig:fig7}, we show again the complete primordial NN contributions,
including isospin mixing and shadowing. But we also show the results
one would obtain by omitting isospin mixing (but considering shadowing),
and by omitting shadowing (but considering isospin mixing). Obviously,
both effect are crucial. In particular the isospin mixing is responsible
for getting a nuclear modification factor of less than unity. %
\begin{figure}
\includegraphics[scale=0.75]{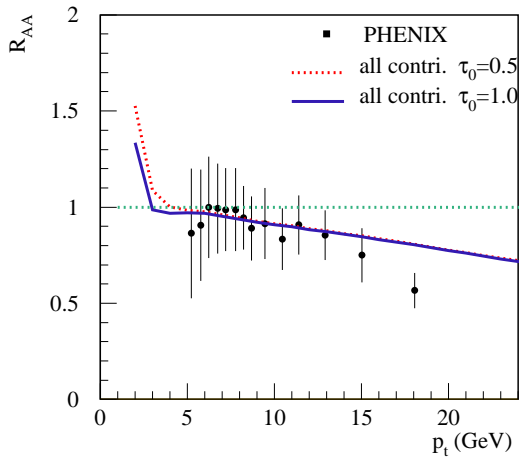}

\caption{\label{fig:fig8} The complete calculation of the nuclear modification
factor is compared to PHENIX data\cite{PHENIX data}. Dotted line:
$\tau_{0}$=0.5~fm/c; solid line: $\tau_{0}$=1~fm/c. Full squares:
the same data as in Fig.\ref{fig:fig6}.}
\end{figure}

Finally we show in Fig.\ref{fig:fig8} the complete calculation, for
two different options of the initial time for the hydrodynamical evolution.
The two curves are almost identical. As mentioned earlier, the total
contribution to the nuclear modification factor is less than unity,
due to isospin correction. However, the experimental data drop systematically
below the theoretical curve, and it seems difficult with the processes
discussed in this paper, to get the theoretical curve further down.
So maybe some \char`\"{}new physics\char`\"{}? Before answering this
question, one should not forget the large uncertainties of the experimental
pp reference, which is a fit function passing between strongly fluctuating
data points at large pt.

\begin{acknowledgments}
This work is supported by the Natural Science Foundation of China
under the project No. 10505010 and by MOE of China under project 
No. IRT0624. F.M.LIU thanks the IN2P3/CNRS and
Subatech for their hospitality during her visit in Nantes. 
\end{acknowledgments}

\end{document}